\documentclass[aps,pra,twocolumn,showpacs,superscriptaddress]{revtex4-2}
\usepackage{amsmath,amssymb,color}
\usepackage[colorlinks=true,urlcolor=blue,citecolor=blue]{hyperref}

\usepackage{graphicx}
\usepackage{dcolumn}
\usepackage{bm}
\usepackage{subfigure}
\usepackage{color}

\usepackage[english]{babel}
\usepackage{graphicx}

\usepackage{mathptmx}
\usepackage{helvet}
\usepackage{courier}
\usepackage[T1]{fontenc}
\usepackage{trajan}
\usepackage{subfigure}
\usepackage{amsmath,amsfonts,amssymb}
\usepackage{graphicx}
\usepackage{epsfig}
\usepackage{extarrows}
\usepackage{mathrsfs}
\usepackage{lineno,hyperref}
\usepackage{color}
\usepackage{amscd,amsbsy,amsgen,amsopn,amstext,amsxtra}
\usepackage{url}
\usepackage{float}

\begin{document}

\title{Feedback-Controlled Magnon-Atom Entanglement and Photon Statistics}

\author{Yiping Lu}
\email{20a0402180@cjlu.edu.cn}
\affiliation{College of Optical and Electronic Technology, China Jiliang University, Hangzhou 310018, China}

\author{Yulong Hu}
\affiliation{College of Optical and Electronic Technology, China Jiliang University, Hangzhou 310018, China}

\author{Zhuzheng Han}
\affiliation{College of Optical and Electronic Technology, China Jiliang University, Hangzhou 310018, China}

\author{Shirong Lin}
\email{shironglin@gbu.edu.cn}
\affiliation{School of Physical Sciences, Great Bay University,Dongguan 523000, China}
\affiliation{Great Bay Institute for Advanced Study, Dongguan 523000, China}

\begin{abstract}
Quantum systems face inherent challenges in achieving precise control, and solving the Schr\"{o}dinger equation is often intractable for complex hybrid platforms. Here, for the first time, we introduce a magnon into a feedback-controlled quantum system. To solve the dynamics numerically and efficiently, we employ a Long Short-Term Memory network, a machine learning approach, to propagate the probability amplitudes to the steady state. By applying coherent feedback, we effectively stabilize the intracavity state. Our results reveal that the photon-photon correlation function and the concurrence, a measure of magnon-atom entanglement, exhibit periodic oscillations with the cavity-mirror distance, and that feedback significantly enhances both antibunching and bunching when the detuning is varied. These findings not only demonstrate the power of artificial intelligence in quantum dynamics simulation, but also open a promising route for on-demand quantum state engineering in hybrid magnonic systems, with potential applications in quantum networks and quantum information processing.
\end{abstract}

\flushbottom
\maketitle
%
%
\thispagestyle{empty}

\section*{\label{sec:level1} Introduction}

The pursuit of advanced quantum technologies has driven the exploration of hybrid systems that coherently couple distinct quantum excitations. Among these, cavity-based platforms integrating magnons (collective spin-wave quanta), photons, and phonons have emerged as versatile testbeds for fundamental physics and quantum information processing \cite{sup1sup, prlShen}. In particular, the realization of strong coupling between magnons and cavity photons \cite{sup1sup}, as well as the recent demonstration of tripartite magnon–photon–phonon strong coupling \cite{shen2025cavity}, highlight the potential of these systems for coherent transduction between optical, microwave, and mechanical domains \cite{sup6sup}. 

Magnonics, the study of spin waves and their quanta, has witnessed remarkable progress owing to its rich phenomenology and compatibility with diverse material platforms. Notable breakthroughs include the observation of magnonic frequency combs via nonlinear scattering \cite{wang2021magnonic}, magnetization switching driven by magnon-mediated spin torque \cite{wang2019magnetization}, and the prediction of topological magnon insulators in two-dimensional ferromagnets \cite{yuan2025prediction}. These advances, together with pioneering work on all-magnonic spin-transfer torque \cite{yan2011all} and magnonic Weyl semimetals \cite{su2017magnonic}, underscore the versatility of magnons as information carriers. Furthermore, the hybridization of magnons with vortices and phonons \cite{li2024magnon, ding2026phonon} offers new routes for manipulating spin dynamics. In addition, strong interlayer magnon-magnon coupling has been realized in magnetic metal-insulator hybrid nanostructures \cite{chen2018strong}, providing another avenue for controlling magnonic interactions. A comprehensive overview of recent developments can be found in the magnonics roadmap \cite{flebus20242024}.

The core physics of a cavity magnomechanical system is governed by three principal coupling pathways. First, magnon–photon coupling arises from the Zeeman interaction between the magnetization of a ferrimagnetic sample (e.g., a yttrium iron garnet sphere) and the microwave cavity field, leading to a coherent beam-splitter interaction that can enter the ultrastrong-coupling regime \cite{sup1sup, huang2026coupling}. Second, photon–phonon coupling is the well-known radiation-pressure interaction in cavity optomechanics, which enables cooling, amplification, and nonlinear dynamics \cite{sup2sup}. Third, in magnetostrictive materials, magnetization and lattice strain are coupled dispersively, providing a direct magnon–phonon interface \cite{ding2026phonon, liu2026rotation}. The interplay of these interactions gives rise to hybrid normal modes and complex quantum correlations.

When all three excitations are confined in a single resonator, the system enters the cavity opto–magno–mechanical regime, where coherent energy exchange and nonclassical correlations can be engineered. The generation and manipulation of tripartite entangled states among photons, phonons, and magnons have been extensively investigated in recent years \cite{Wang2025OPA, Lu2025Barnett, Jiao2025Phase}. Specifically, phase-controlled robust tripartite entanglement has been achieved by tuning the phase difference of two coherent driving fields in opposite input directions \cite{Jiao2025Phase}, while synergistic optical and mechanical parametric amplifications have been shown to significantly enhance tripartite photon-phonon-magnon entanglement \cite{Wang2025OPA}. Moreover, nonreciprocal bipartite and tripartite entanglement among magnons, photons, and phonons has been realized via the Barnett effect in a rotating Yttrium Iron Garnet(YIG) sphere \cite{Lu2025Barnett}. For instance, the statistical properties of cavity photons, such as antibunching and bunching, have been linked to the entanglement of atomic ensembles \cite{SongEntangleCreat, wang2025}. In hybrid systems, simultaneous blockade of photons, magnons, and phonons has been predicted via tailored tripartite nonlinearities \cite{2, xie2022nonreciprocal}. Recent work has also demonstrated magnon blockade in spin-magnon systems with frequency detuning \cite{Zhao2025MagnonBlockade}, and nonreciprocal unconventional magnon blockade via the Sagnac-Fizeau shift in optomagnonic systems \cite{Deng2025Nonreciprocal}. Continuous-variable entanglement generation has been extensively investigated \cite{SciRep3833, PhysRevApplied044048, Peng2025EntanglementSteering, Kumar2025ManyBody, Chen2025Levitated}. Moreover, the control of photon blockade and the transition between conventional and unconventional blockade regimes have been studied in various configurations \cite{PhysRevA013705, PhysRevA023856, PhysRevA053703}. Coherent manipulation via external driving or waveguide coupling has also been explored to tailor quantum fluctuations \cite{OptExpress32967, AnnPhys00606, OptExpress25901}. Additionally, entanglement characteristics of levitated magnomechanical systems have been studied, revealing that bipartite and tripartite entanglements can be manipulated by changing the radius and oscillation frequency of the levitated sphere \cite{Chen2025Levitated}.

To actively stabilize and enhance these quantum properties, real-time feedback control has emerged as a powerful technique \cite{sup5sup, sup7sup}. By continuously monitoring the output field and feeding back a processed signal, one can cool mechanical modes, suppress noise, generate steady-state entanglement \cite{sup3sup}, or induce nonreciprocal interactions \cite{sup4sup}. For example, time-delayed coherent feedback in cavity Quantum Electrodynamics has been shown to dramatically enhance or suppress photon antibunching and bunching by reinforcing specific entangled states \cite{three}. More recently, coherent feedback loops have been applied to optomechanical systems for high-fidelity state transfer \cite{feedbackAmazioug2020}, to magnomechanics for nonreciprocal entanglement via the Barnett effect \cite{feedbackAmaziougBarnett}, and to enhance tripartite magnon–photon–phonon entanglement. Furthermore, Barnett-effect-induced rotation enables controllable nonreciprocal magnon and photon blockade \cite{PhotonBlockade}, while strong antibunching has been realized in double-cavity optomechanical systems with squeezed light \cite{PhotonBlockadeAntibunching}. Recent theoretical advances have also explored nonreciprocal macroscopic entanglement through magnon squeezing in cavity magnomechanics \cite{Amazioug2026MagnonSqueezing}, and nonreciprocal steering between optical and microwave waves via Bogoliubov cooling in cavity optomagnonic systems \cite{Kong2025NonreciprocalSteering}.

The inclusion of time-delayed feedback, however, complicates the theoretical treatment, as the system dynamics becomes non-Markovian and analytically intractable. Advanced computational methods, particularly from machine learning, have proven effective in solving such complex quantum dynamics. Reservoir computing has been adapted to solve the time-dependent Schrödinger equation for high-dimensional systems \cite{RCforE}, and Long Short-Term Memory (LSTM) networks have demonstrated exceptional capability in capturing nonlinear wave dynamics \cite{LSTMbased, LSTMintrodu1, LSTMintrodu2, lstmReview}. These data-driven approaches offer promising avenues for simulating and optimizing feedback-controlled quantum platforms.

Parallel to these developments, efficient quantum state characterization tools have been proposed, including PhaseLift-based tomography and direct fidelity estimation \cite{lu2015quantum, lu2016minimum, lu2017enhancing, lu2020minimum}. Breakthroughs in super-resolution imaging have also emerged, combining neural networks with vectorial Debye integrals to achieve sub-diffraction focusing \cite{jin2025surpassing, jin2026endtoend}. In the realm of light–matter interactions, photonic platforms have served as a bridge connecting classical and quantum simulations of physical phenomena \cite{lin2022all, lin2023controllable, yang2025chiral}. All-optical spin control with magnetic textures generation has been demonstrated \cite{zhang2026skyrmion, lin2019all, qifan2026magnetic, adamantopoulos2024spin}. In addition, nonreciprocal amplification \cite{zhao2022nonreciprocal}, wideband microwave conversion via magnon nonlinearity \cite{wu2024wideband}, and tunable slow-to-fast light conversion through optomagnonic coupling \cite{liu2026rotation} further enrich the toolbox for hybrid quantum systems. Recent studies have also investigated the dynamics of quantum entanglement between photon and phonon modes in Coulomb-coupled optomechanical cavity magnonic systems \cite{Mikki2026Coulomb}, and proposed protocols for remote magnon-phonon entanglement in waveguide magnomechanics \cite{Qi2026Remote} as well as nonreciprocal quantum coherence via the Barnett effect \cite{Lu2026NonreciprocalCoherence}.

In this work, we introduce, for the first time, a magnon degree of freedom into a feedback-controlled hybrid quantum system, and we employ a LSTM network—a machine learning approach—to numerically solve the equation of motion until the steady state is reached. We present detailed results on the controlled photon statistics and entanglement in such a hybrid cavity magnomechanical system, demonstrating the synergy between feedback control and modern computational techniques. The integration of magnonic, photonic, and phononic degrees of freedom in cavity systems, augmented by such approaches, constitutes a vibrant frontier in quantum engineering. The following sections describe the model, the feedback scheme, and the numerical findings.

\section*{\label{sec:level1}Long Short-Term Memory}

In this section, we introduce the LSTM method. LSTM is a special type of recurrent neural network designed to address the vanishing and exploding gradient problems of traditional Recurrent Neural Networks (RNNs). It introduces a cell state and gating mechanisms to selectively retain or forget information, thereby enabling effective learning of long-range dependencies. Compared with traditional RNNs, LSTMs are better at capturing long-distance relationships in sequences. LSTMs also offer great flexibility, as their gating mechanisms allow dynamic adjustment of the information flow to accommodate different task requirements.

\begin{figure}[!h]
\begin{center}
\includegraphics[width=1\linewidth]{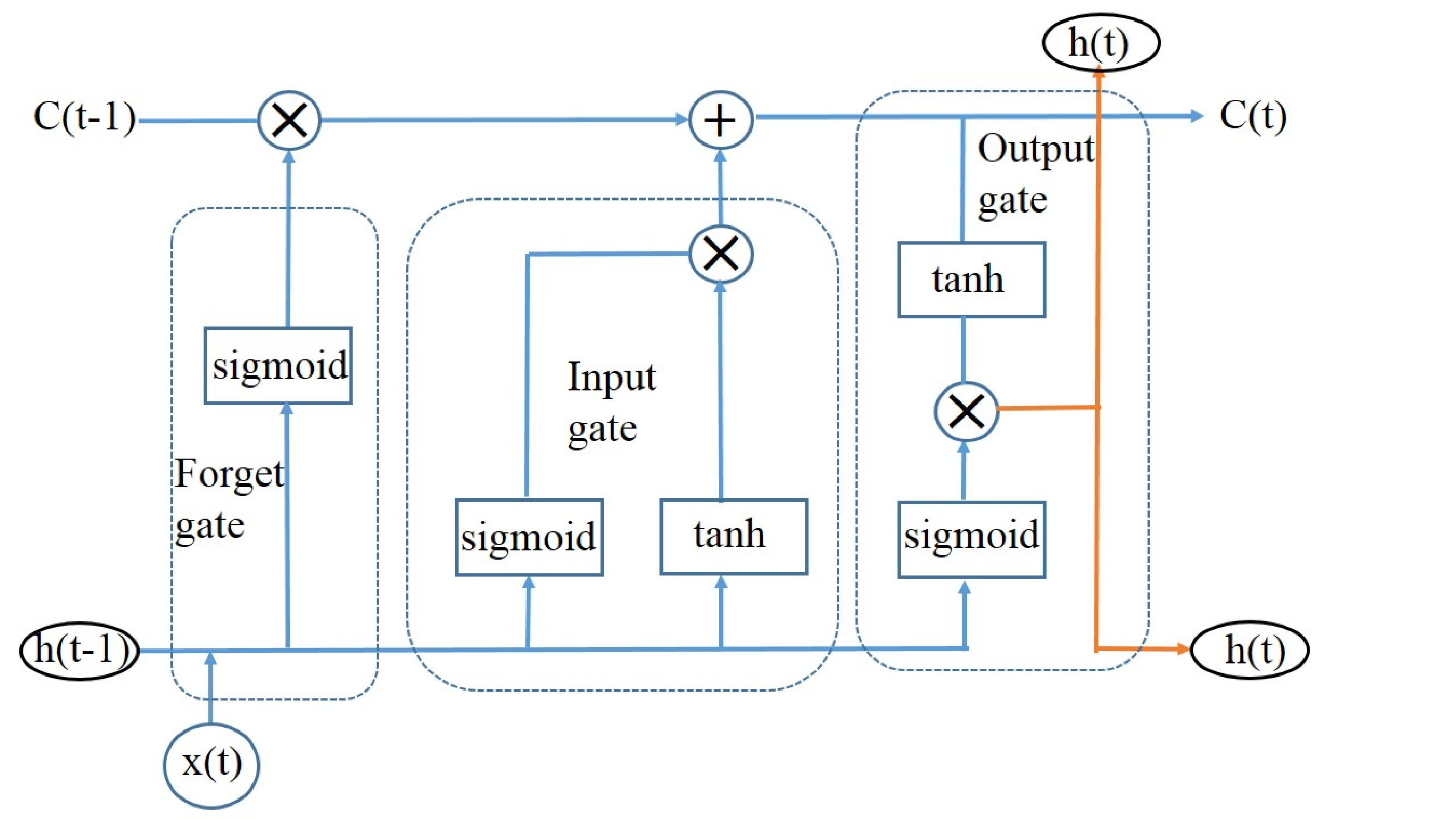}\\
\caption{Structure of an LSTM unit. Here, $C(t)$ is the cell state vector at time step $t$, $C(t-1)$ is the cell state from the previous time step, $h(t)$ is the hidden state vector at time step $t$, $h(t-1)$ is the hidden state from the previous time step, and $x(t)$ is the input vector at the current time step $t$. The sigmoid activation function outputs values between 0 and 1, determining how much information to pass through. The $\tanh$ activation function outputs values between $-1$ and $1$, regularizing candidate values. $\times$ denotes point-wise multiplication (Hadamard product), and $+$ denotes point-wise addition.}\label{LSTMjiegoutu}
\end{center}
\end{figure}

A standard LSTM unit consists of an input gate, a forget gate, an output gate, and a cell state \cite{LSTMintrodu1,LSTMintrodu2}, as shown in Fig.~\ref{LSTMjiegoutu}. Its core idea is to regulate the flow of information through gating mechanisms.

Specifically, the first gate is the forget gate, which decides what information to discard from the cell state (memory unit). Since the sigmoid function outputs a value between 0 and 1, it is used to determine what to forget. Its mathematical expression is
\begin{eqnarray}
f_t = \sigma(W_f \cdot [h(t-1), x(t)] + b_f),
\end{eqnarray}
where $\sigma$ is the sigmoid function \cite{lstmReview}. This gate decides which information to discard from the cell state: 0 means completely forget and 1 means completely retain. $W_f$ and $b_f$ denote the weight matrix and the bias term for the forget gate, respectively, and are parameters learned by the model. $h(t-1)$ represents the hidden state from the previous time step (the output of the previous unit), and $x(t)$ is the input at the current time step. $[h(t-1), x(t)]$ denotes the concatenation of these two vectors.

The second gate is the input gate, which controls the addition of new information \cite{lstmReview}. Together with the candidate update, it updates the memory unit. Its mathematical expression is
\begin{eqnarray}
i_t = \sigma(W_i \cdot [h(t-1), x(t)] + b_i).
\end{eqnarray}
This gate decides what new information to store. $W_i$ and $b_i$ denote the weight matrix and the bias term for the input gate, respectively, and are parameters to be learned. The meanings of $h(t-1)$, $x(t)$, and $[h(t-1), x(t)]$ are the same as in the forget gate. In addition, we use $\tilde{C}(t)$ to generate new candidate information, which is activated by the $\tanh$ function. The candidate expression is
\begin{eqnarray}
\tilde{C}(t) = \tanh(W_C \cdot [h(t-1), x(t)] + b_C),
\end{eqnarray}
where $\tanh$ is the hyperbolic tangent function, which maps the values to the interval $[-1, 1]$ \cite{lstmReview}. This step determines what candidate information to incorporate. $W_C$ and $b_C$ denote the weight matrix and the bias term for generating new candidate information, respectively, and are parameters to be learned. Combining these operations, the cell state is updated as
\begin{eqnarray}
C(t) = f_t \odot C(t-1) + i_t \odot \tilde{C}(t).
\end{eqnarray}
Here, we perform element-wise multiplication of $f_t$ with $C(t-1)$, and element-wise multiplication of $i_t$ with $\tilde{C}(t)$, then sum the two parts \cite{lstmReview}.

The last gate is the output gate, which determines the output based on the current cell state \cite{lstmReview}. When facing specific tasks, it selects useful information from the memory. The mathematical expression of the output gate is
\begin{eqnarray}
o(t) = \sigma(W_o \cdot [h(t-1), x(t)] + b_o).
\end{eqnarray}
This gate decides which information to output. $W_o$ and $b_o$ denote the weight matrix and the bias term for the output gate, respectively, and are parameters to be learned.

Based on $C(t)$ and $o(t)$, the hidden state $h(t)$ is obtained. Specifically, the cell state $C(t)$ is first passed through a $\tanh$ function, which maps its values to the range $[-1, 1]$, and then multiplied element-wise with the output gate $o(t)$. This yields the hidden state to be passed to the next LSTM unit. Therefore, the mathematical expression for the hidden state is
\begin{eqnarray}
h(t) = o(t) \odot \tanh(C(t)).
\end{eqnarray}

Combining the forget gate, the input gate, and the output gate forms a cell \cite{lstmReview}. The cell acts as an information highway across time steps, allowing linear information transmission through the gating mechanism. During computation, the gradient along the cell state experiences almost no attenuation, thus avoiding the vanishing gradient problem. At the same time, the sigmoid function controls the information flow to prevent exploding gradients. This effectively preserves long-term memory.

\section*{\label{sec:level1}Model}

\begin{figure}[!h]
\begin{center}
\includegraphics[width=1\linewidth]{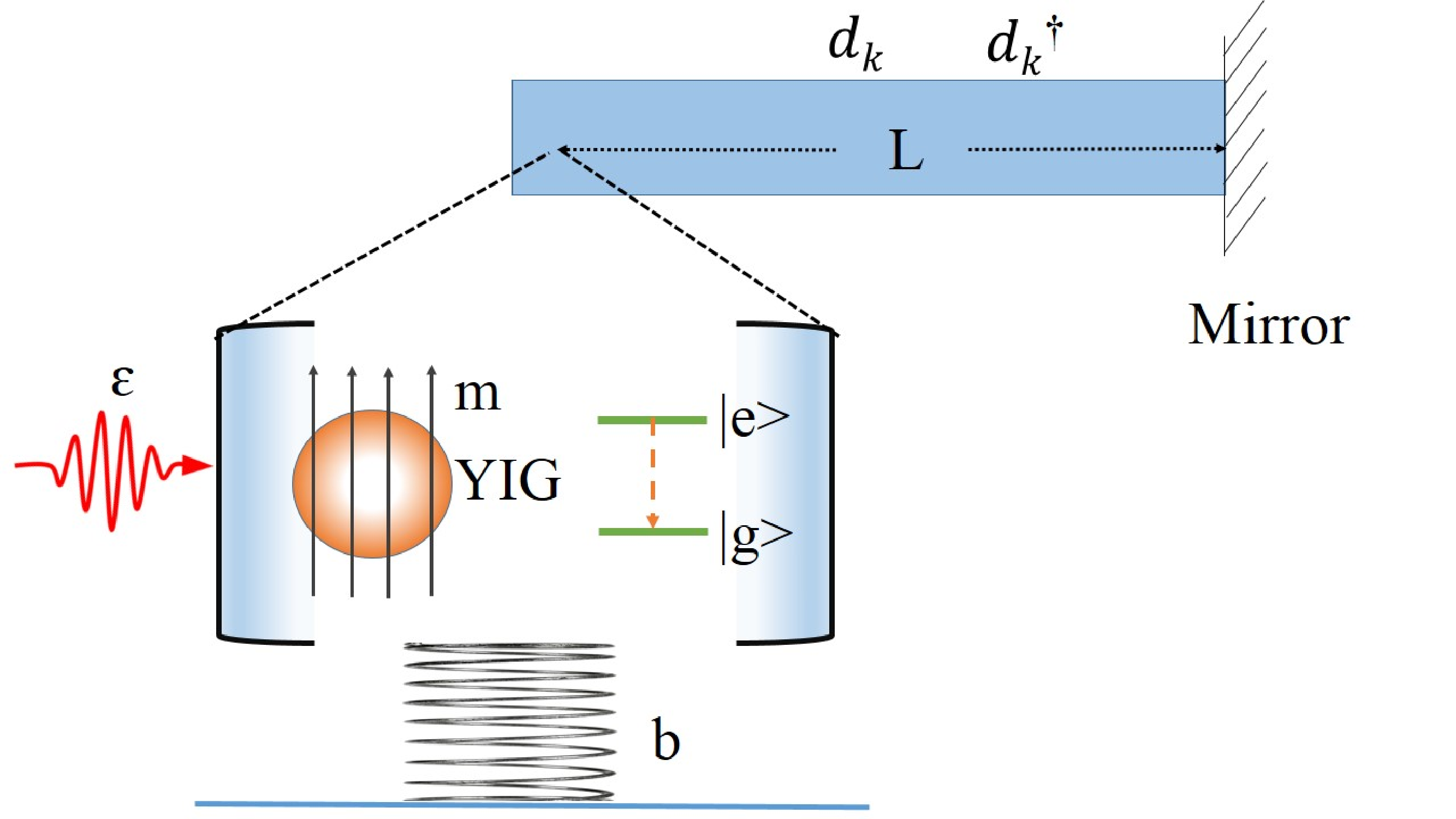}\\
\caption{Realization of quantum feedback via a photonic waveguide. A magnon and a two-level emitter trapped in a small cavity interact with the photons inside the cavity. The driving field is weakly coupled to the cavity mode with frequency $\omega_l$. Part of the cavity loss is fed back into the cavity from the waveguide at a distance $L$ from the cavity. Other losses are fed back into the cavity via the phonon, denoted by $b$ and $b^\dagger$.}\label{shiYiTu}
\end{center}
\end{figure}

In this section, we introduce the model of the system. A magnon and a two-level atom are trapped in an optical cavity. Part of the output signal is fed back into the system via a mechanical spring, while another part is fed back via a mirror embedded in a waveguide. The distance between the cavity and the mirror is denoted by $L$. The annihilation and creation operators for photons in the waveguide mode (labeled by wave vector $k$) are denoted by $\hat{d}_k$ and $\hat{d}_k^\dagger$, respectively. The magnon mode is described by the operators $\hat{m}$ and $\hat{m}^\dagger$. The two-level atom is characterized by its excited state $|e\rangle$ and ground state $|g\rangle$. Furthermore, the cavity field is coupled to a mechanical oscillator (modeled as a spring), whose annihilation and creation operators are denoted by $\hat{b}$ and $\hat{b}^\dagger$.

The Hamiltonian for the photon mode and the magnon in the cavity is written as $H_{am}$,
\begin{eqnarray}
H_{am}=\omega_c a^{\dagger} a+\omega_m m^{\dagger} m+G_m (a^{\dagger}m + a m^{\dagger}).
\end{eqnarray}
Here, $\omega_c$ is the cavity frequency, $a$ is the photon annihilation operator, $a^{\dagger}$ is the photon creation operator, $\omega_m$ is the magnon frequency, and $G_m$ is the strength of the photon–magnon interaction. Considering the feedback from the external continuum, the Hamiltonian for the photon mode in the cavity and the waveguide is written as $H_{ad}$,
\begin{eqnarray}
H_{ad}= \omega_d d_k^\dagger d_k +\int dk \bigl[G(k,t) a^\dagger d_k + G^*(k,t) d_k^\dagger a \bigr].
\end{eqnarray}
Here, $\omega_d$ is the frequency of the photon mode in the waveguide, and $G(k,t)$ is the external coupling amplitude, defined as
\begin{eqnarray}
G(k,t)=G_0\sin(k L)\exp[i(\omega_l-\omega_k)t].
\end{eqnarray}
Here, $G_0$ denotes the bare tunnel coupling strength between cavity photons and external modes in the continuum. $\omega_l$ corresponds to the external driving frequency, $\omega_k$ is the frequency of the waveguide mode with wave number $k$, and $\omega_k=c|k|$ with $c$ being the speed of light in the waveguide.

In the system, the two-level-system lowering operator is $\sigma$, defined as $\sigma=|g\rangle\langle e|$, and $\sigma^\dagger$ is the raising operator. Therefore, the Hamiltonian for the photon mode and the two-level emitter in the cavity is written as $H_{\sigma a}$,
\begin{eqnarray}
H_{\sigma a}= \omega_a \sigma^\dagger \sigma+ g_a(\sigma a^\dagger+ \sigma^\dagger a),
\end{eqnarray}
where $\omega_a$ is the resonance frequency of the two-level system, and $g_a$ is its coupling strength to the cavity mode. In addition, the Hamiltonian for the mechanical oscillator and its interaction with the photon mode is written as $H_{b a}$,
\begin{eqnarray}
H_{b a}= \omega_b b^\dagger b+ g_b(b a^\dagger+ b^\dagger a),
\end{eqnarray}
where $\omega_b$ is the phonon frequency, $b^\dagger$ is the phonon creation operator, and $b$ is the annihilation operator. The phonon is coupled to the photon mode with strength $g_b$. Furthermore, the Hamiltonian for the driving of the photon mode is written as $H_{\epsilon}$,
\begin{eqnarray}
H_{\epsilon}= \epsilon (a^\dagger e^{i\omega_l t}+a e^{-i\omega_l t}) \label{Hepsilon},
\end{eqnarray}
where the driving strength is represented by $\epsilon$, and $\omega_l$ is the driving frequency.

Adding all these Hamiltonians together, we obtain
\begin{eqnarray}
H=H_{am}+H_{ad}+H_{\sigma a}+H_{ba}+H_{\epsilon}.
\end{eqnarray}

Define $a_{\pm}=\frac{1}{\sqrt{2}}(a\pm m)$ and perform the rotating-frame approximation. The Hamiltonian $H$ is written as
\begin{equation}
\begin{aligned}
H_{\text{eff}} &= \Delta a_+^\dagger a_+ + (\Delta-2G_m) a_-^\dagger a_- + \omega_b b^\dagger b + \Delta_a \sigma^\dagger\sigma \\
&\quad - \eta (a_+^\dagger a_- b + a_+ a_-^\dagger b^\dagger) + \eta_a (a_+^\dagger \sigma + a_+\sigma^\dagger) \\
&\quad + \int dk \Big[ \delta_k d_k^\dagger d_k + \bigl( G(k,t) a_+^\dagger d_k + G^*(k,t) d_k^\dagger a_+ \bigr) \Big] \\
&\quad + \Omega_e (\sigma+\sigma^\dagger),
\end{aligned}
\end{equation}
where the detunings are $\Delta=\delta+G_m$, $\delta=\omega_m-\omega_l$. Here, $\eta$ is the effective photon–magnon–phonon three-wave mixing coupling strength, and $\eta_a$ is the effective photon–magnon–atom three-wave mixing coupling strength. The detuning from the atomic resonance is $\Delta_a=\omega_a-\omega_l$, and $\Omega_e$ is the effective driving strength derived from Eq.~\ref{Hepsilon}. The detuning from the resonance of the photon mode in the waveguide is $\delta_k=\omega_d-\omega_l$.

Since we consider the weak-driving limit, the state of the system is truncated to the few-excitation subspace. Therefore, the wave function can be approximately written as
\begin{equation}
\begin{aligned}
|\psi(t)\rangle &= C_{g0000}|g,0_+,0_-,0_b,0\rangle \\
&\quad + C_{g1000}|g,1_+,0_-,0_b,0\rangle + C_{g0110}|g,0_+,1_-,1_b,0\rangle\\
&\quad + C_{e0000}|e,0_+,0_-,0_b,0\rangle + C_{g2000}|g,2_+,0_-,0_b,0\rangle\\
&\quad + C_{g1110}|g,1_+,1_-,1_b,0\rangle + C_{e1000}|e,1_+,0_-,0_b,0\rangle \\
&\quad + C_{g0220}|g,0_+,2_-,2_b,0\rangle + C_{e0110}|e,0_+,1_-,1_b,0\rangle \\
&\quad + \int dk\, C_{g000k}|g,0_+,0_-,0_b,k\rangle \\
&\quad + \int dk\, C_{g011k}|g,0_+,1_-,1_b,k\rangle \\
&\quad + \int dk\, C_{g100k}|g,1_+,0_-,0_b,k\rangle \\
&\quad + \int dk\, C_{e000k}|e,0_+,0_-,0_b,k\rangle \\
&\quad + \iint dk\,dk'\, C_{g000kk'}|g,0_+,0_-,0_b,k,k'\rangle.
\end{aligned}
\end{equation}
Here, we choose the basis $|j,n_+,n_-,j_b,k\rangle$, where $j$ describes the two-level emitter ($g$ for ground state, $e$ for excited state), $n_+$ is the occupation number of the supermode $(a+m)/\sqrt{2}$, $n_-$ is that of $(a-m)/\sqrt{2}$, $j_b$ is the phonon number, and $k,k'$ label external photon modes. Correspondingly, the coefficient $C_{j,n_+,n_-,j_b,k}$ denotes the probability amplitude of the state.

To calculate the evolution of these states, the dynamics is governed by the Schrödinger equation:
\begin{eqnarray}
i \frac{\partial}{\partial t}|\varphi(t)\rangle = H_{\text{eff}} |\varphi(t)\rangle.
\end{eqnarray}
Since weak driving is considered, the coefficient $|C_{g0000}|$ is approximated as unity, so that the equations of motion are obtained (see Appendix 1). The equation of motion [Eq.~(\ref{EqMotion})] is solved numerically by two approaches: one is the iteration method, and the other is LSTM. Since the number of equations in Eq.~(\ref{EqMotion}) is large, it is slow to obtain the dynamics using the iteration method. We run the iteration method for a certain number of steps and then switch to LSTM to predict the future values of the coefficients until the steady states are obtained. To obtain the steady state directly without knowing the evolution process, the derivatives of the probability amplitudes in Eq.~(\ref{EqMotion}) are set to zero. Thus, we have Eq.~(\ref{EqMotionSteady}). The traditional method is applied to solve this linear equation, as shown in Appendix 2.

Once the steady-state solution of the probability amplitudes $C_{j,n_+,n_-,j_b,k}$ is obtained, the probabilities $P_{j,n_+,n_-,j_b,k}$ are calculated using
\begin{eqnarray}
P_{j,n_+,n_-,j_b,k}=|C_{j,n_+,n_-,j_b,k}|^2. \label{pDefine}
\end{eqnarray}

Furthermore, based on the steady state, the photon–photon correlation function (usually the second-order correlation function $g^{(2)}(\tau)$) quantifies the probability of detecting a second photon at time $\tau$ after a first photon, relative to a random (coherent) light source. Its zero-delay value $g^{(2)}(0)$ is key:
\begin{itemize}
\item $g^{(2)}(0)<1$: antibunching – photons arrive regularly, a hallmark of non-classical light.
\item $g^{(2)}(0)=1$: coherent light (e.g., a laser).
\item $g^{(2)}(0)>1$: bunching – classical or thermal light.
\end{itemize}
$g^{(2)}(0)$ is the primary experimental observable used to detect both conventional and unconventional photon blockade, because both mechanisms aim to suppress multi-photon events, resulting in strong antibunching with $g^{(2)}(0)<1$. We calculate $g^{(2)}(0)$ using
\begin{eqnarray}
g^{(2)}(0)=\frac{\langle \varphi(t)| a^\dagger a^\dagger a a |\varphi(t)\rangle}{\langle \varphi(t)|a^\dagger a |\varphi(t)\rangle^2}. \label{g2Define}
\end{eqnarray}
where $\langle \varphi(t)|a^\dagger a |\varphi(t)\rangle$ is the mean intracavity photon number. However, $g^{(2)}(0)$ alone cannot distinguish between the two types; additional information is required.

For conventional photon blockade, the transition frequency between $|0\rangle$ and $|1\rangle$ is not equal to the energy gap between $|1\rangle$ and $|2\rangle$. When the first photon enters the system, the state is excited to $|1\rangle$; this makes it difficult for the second photon to enter, because the transition frequency required between $|1\rangle$ and $|2\rangle$ differs from that between $|0\rangle$ and $|1\rangle$. When the driving frequency is tuned to $\omega_{01}$, it excites the first photon with high efficiency. However, once a photon exists inside the cavity, it requires the frequency $\omega_{12}$ to be continuously excited. The key condition for realizing photon blockade is that the coupling strength be much larger than the loss rate $\kappa$.

In our model, although the coupling strength is large, the photon blockade arises mainly from interference caused by the feedback from the waveguide with the mirror, and the driving is weak as well. Therefore, it is an unconventional photon blockade.

\section*{\label{sec:level1} Results}

We present the dependence of the observables on several key parameters: the distance $L$ between the mirror and the cavity, the detuning $\Delta$, the external drive $\Omega_e$, the magnon–photon coupling $\eta$, and the feedback strength $G_0$. The default parameters (unless varied) are: $\kappa = 5\times10^{-6}$ (normalized units), $\eta = 5\kappa$, $\eta_a = (6/\sqrt{2})\kappa$, $\Omega_e = 0.1\kappa$, $G_m = 200\kappa$, $\delta = 2\sqrt{2}\eta$, $\Delta = \delta+G_m$, $\omega_b = 2G_m$, $\gamma=\kappa$, $c=299.7925$ (mm/ns), $\omega_0 = 1.1\times10^5\kappa$, $g=40\kappa$, and $\omega_l = \Delta+\omega_0$. The feedback distance is fixed at $L = \tau c/2$ with $\tau=2\pi/\omega_0$, and $G_0 = \sqrt{2c\kappa/\pi}$ unless scanned.

\begin{figure}[!h]
\begin{center}
\includegraphics[width=1\linewidth]{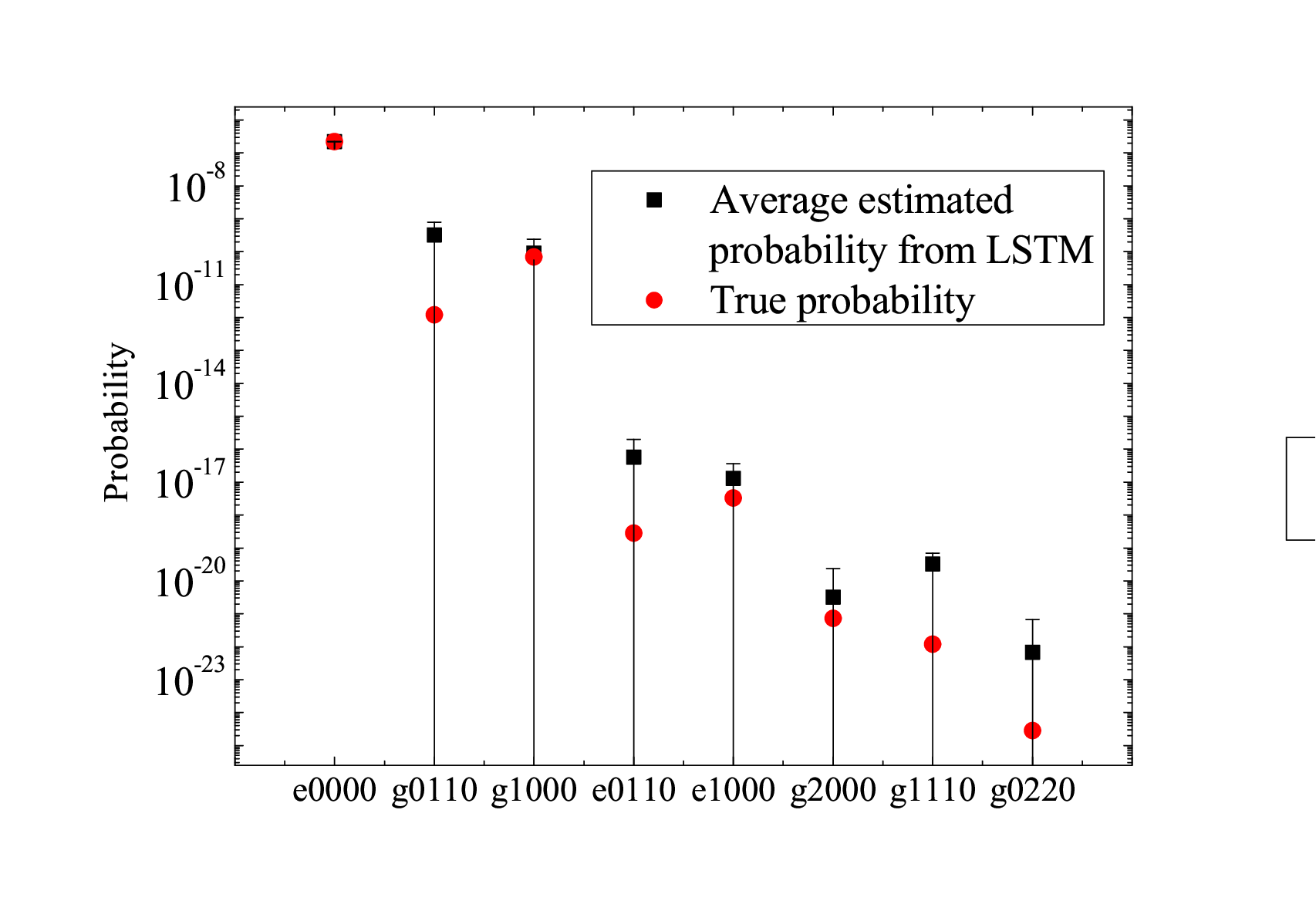}\\
\caption{Probabilities predicted by the LSTM. The average value is calculated from ten predictions; the mean square error is denoted by the error bar. The true value is denoted by the red point.}\label{ProbaStablize}
\end{center}
\end{figure}

Using the LSTM and Eq.~(\ref{pDefine}), the probabilities for different states are obtained and listed in Fig.~\ref{ProbaStablize}. Due to the inherent stochasticity or different seeds applied in the LSTM, each prediction run yields a slightly different value. Therefore, we perform ten independent prediction runs and take the median of these ten values as the final prediction. As shown in Fig.~\ref{ProbaStablize}, the probabilities of the single-excited states $|g1000\rangle$, $|g0110\rangle$, and $|e0000\rangle$ are around $10^{-8}$, which is much larger than the values around $10^{-19}$ for the double-excited states $|g2000\rangle$, $|g1110\rangle$, $|g0220\rangle$, $|e1000\rangle$, and $|e0110\rangle$.

Based on the different probabilities for different states, we study the impact of coherent feedback on quantum statistics and entanglement. Specifically, the series of figures (Fig.~\ref{tau} and Fig.~\ref{Delta}) present a coherent investigation into how time-delayed coherent feedback, implemented via a distant mirror in a waveguide, is employed to manipulate quantum statistical properties and entanglement in a hybrid magnon–photon–phonon–atom system.

\subsection{Effect of feedback distance $L$}

The central control knob is the feedback delay, determined by the distance $L$ between the cavity and the mirror ($L = \tau c/2$, where $\tau$ is the round-trip time). This delay and the associated phase shift allow selective reinforcement or suppression of specific quantum pathways within the cavity.

\begin{figure}[!h]
\begin{center}
\includegraphics[width=1\linewidth]{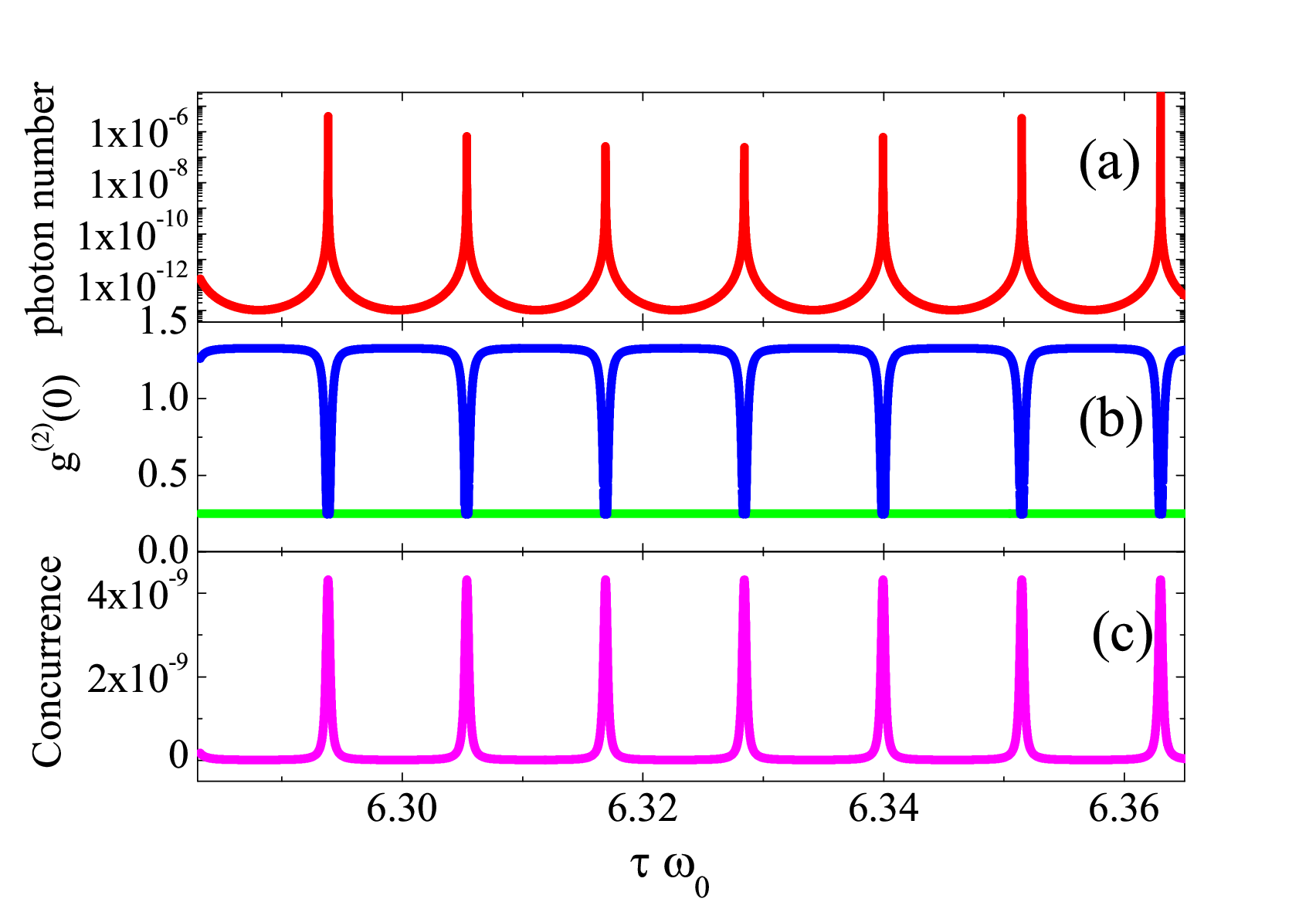}
\caption{(a) Impact of feedback on the photon number. The photon number fluctuates periodically with increasing distance between the cavity and mirror. (b) Impact of feedback on $g^{(2)}(0)$. The $g^{(2)}(0)$ function fluctuates periodically with increasing distance $L$. The case with feedback is denoted by the blue curve; the case without feedback is represented by the green curve. (c) Impact of feedback on the concurrence. The concurrence fluctuates periodically with increasing $L$. The parameters are chosen as follows: $\kappa=5\times10^{-6}\times 0.4\pi\,\text{THz}=2\pi\times 1\,\text{MHz}$, $\eta=5\kappa=10\pi\times 1\,\text{MHz}$, $\eta_a=\frac{6}{\sqrt{2}}\kappa=2\pi\times 4.24\,\text{MHz}$, $\Omega_e=0.1\kappa=2\pi\times 0.1\,\text{MHz}$, $G_m=200\kappa=2\pi\times 200\,\text{MHz}$, $\delta=2\sqrt{2}\eta=2\pi\times 10\sqrt{2}\,\text{MHz}$, $\Delta=\delta+G_m=2\pi\times 214.142\,\text{MHz}$, $\Delta_a=\Delta$, $\omega_b=2G_m=2\pi\times 400\,\text{MHz}$, $\gamma=\kappa=2\pi\times 1\,\text{MHz}$, $c=299.7925\,\text{mm/ns}$, $\omega_0=1.1\times10^5\gamma=2\pi\times 110\,\text{GHz}$, $g=40\gamma=2\pi\times 40\,\text{MHz}$, $\omega_l=\Delta+\omega_0$, $L = \tau c/2 = 1.3627\,\text{mm}$, and $G_0 = \sqrt{2c\gamma/\pi} = 1.095065\times10^9\,\sqrt{\text{mm}}/\text{ns}$.}\label{tau}
\end{center}
\end{figure}

We compute the intracavity photon number, as shown in Fig.~\ref{tau}(a). The mean intracavity photon number $\langle a^\dagger a \rangle$ also exhibits periodic oscillations with $L$. Specifically, feedback modifies the effective damping (or amplification) of the cavity mode. Distances that lead to constructive interference at the cavity frequency enhance the field buildup (higher photon number), while destructive interference suppresses it. This demonstrates the role of feedback in coherent energy control.

Using the same parameters as for the intracavity photon number, we employ the second-order correlation function $g^{(2)}(0)$ to investigate photon statistics as a function of $L$. When feedback is introduced, we calculate $g^{(2)}(0)$ for cavity photons [Eq.~(\ref{g2Define})] as a function of the distance $L$. The result exhibits periodic oscillations with $L$, as shown by the blue curve in Fig.~\ref{tau}(b). This is a direct signature of coherent quantum interference. Specifically, certain distances (phases) constructively interfere with processes that promote two-photon states (bunching, $g^{(2)}(0)>1$), while others destructively interfere with them, favoring single-photon states (antibunching, $g^{(2)}(0)<1$). The feedback loop thus acts as an external, tunable ``interferometer'' to actively switch between classical (bunched) and quantum (antibunched) light statistics from the same system.

Using the same parameter set, we compute the concurrence, as shown in Fig.~\ref{tau}(c). The concurrence (a measure of atom–magnon entanglement) oscillates with the same period as $g^{(2)}(0)$ and the photon number. This is the most critical result, directly linking the external feedback parameter ($L$) to internal quantum correlations. The oscillations confirm that the feedback not only affects light statistics but also actively manipulates the entanglement between the atom and the magnon. Peaks in concurrence coincide with specific feedback phases that reinforce the entangled eigenstates of the hybrid system.

In short, these results demonstrate the periodic modulation of key quantum observables by varying the feedback distance $L$. The synchronized oscillations provide compelling evidence that a single external parameter ($L$) simultaneously and coherently controls: quantum statistics of light ($g^{(2)}$), field amplitude (photon number), and atom–magnon entanglement (concurrence). This validates the promise of coherent feedback as a multifunctional quantum control tool.

\subsection{Effect of detuning $\Delta$}

\begin{figure}[!h]
\begin{center}
\includegraphics[width=1\linewidth]{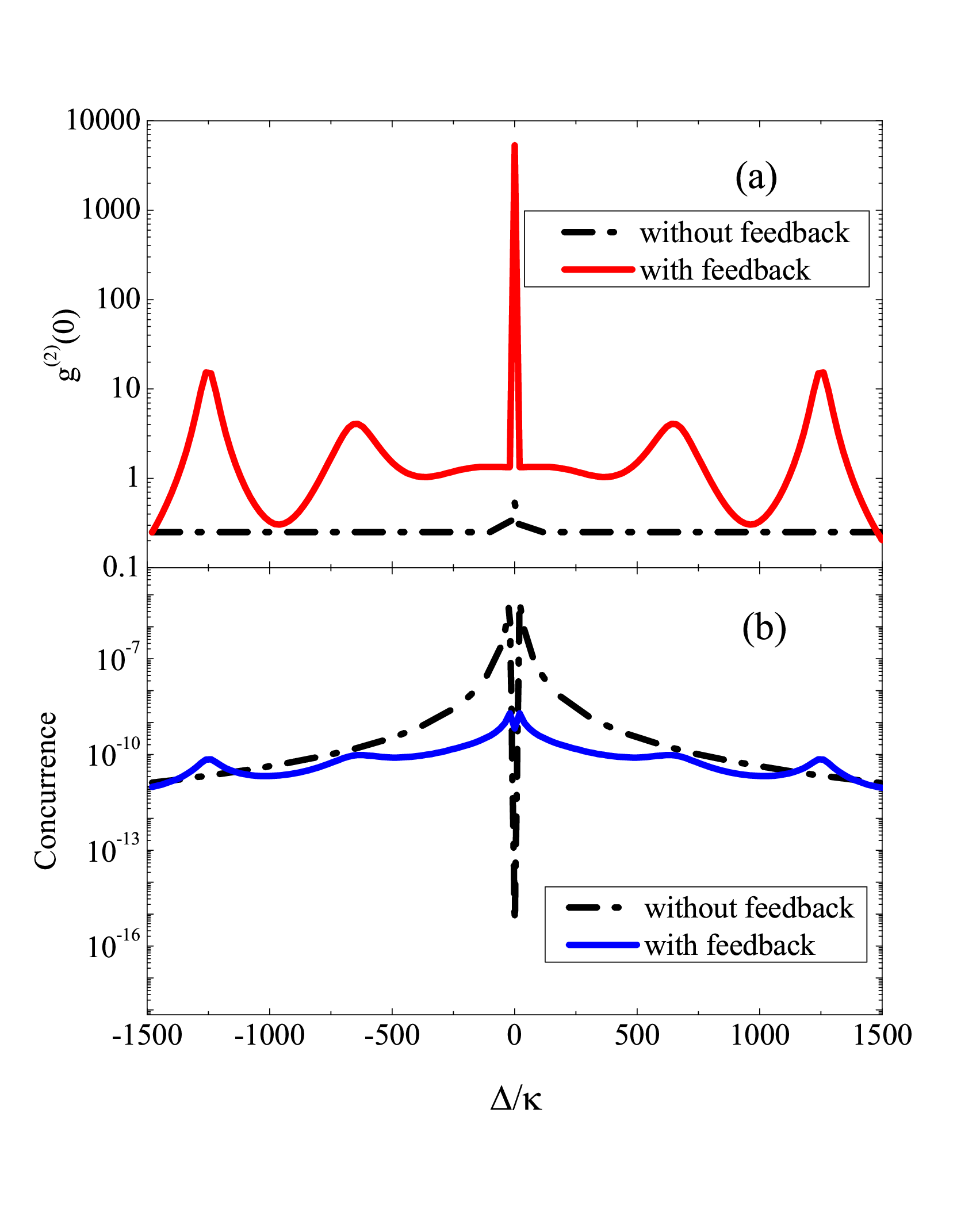}
\caption{(a) Dependence of $g^{(2)}(0)$ on $\Delta$ at a fixed distance $L$ between the cavity and mirror ($\tau=2\pi/\omega_0=0.009\,\text{ns}$, $L = \tau c/2 = 1.3627\,\text{mm}$). (b) Behavior of the concurrence with a fixed distance for increasing $\Delta$ ($\tau=2\pi/\omega_0=0.009\,\text{ns}$, $L = \tau c/2 = 1.3627\,\text{mm}$). As a reference, the case without feedback is shown as a black dash-dotted line ($\tau=0$, $L=0$). The parameters are chosen as follows: $\kappa=5\times10^{-6}\times 0.4\pi\,\text{THz}=2\pi\times 1\,\text{MHz}$, $\eta=5\kappa=10\pi\times 1\,\text{MHz}$, $\eta_a=\frac{6}{\sqrt{2}}\kappa=2\pi\times 4.24\,\text{MHz}$, $\Omega_e=0.1\kappa=2\pi\times 0.1\,\text{MHz}$, $G_m=200\kappa=2\pi\times 200\,\text{MHz}$, $\delta=2\sqrt{2}\eta=2\pi\times 10\sqrt{2}\,\text{MHz}$, $\Delta_a=\Delta$, $\omega_b=2G_m=2\pi\times 400\,\text{MHz}$, $\gamma=\kappa=2\pi\times 1\,\text{MHz}$, $c=299.7925\,\text{mm/ns}$, $\omega_0=1.1\times10^5\gamma=2\pi\times 110\,\text{GHz}$, $g=40\gamma=2\pi\times 40\,\text{MHz}$, $\omega_l=\Delta+\omega_0$, and $G_0 = \sqrt{2c\gamma/\pi}=1.095065\times10^9\,\sqrt{\text{mm}}/\text{ns}$.}\label{Delta}
\end{center}
\end{figure}

With the feedback distance $L$ fixed at an optimal value, we scan the system detuning $\Delta$. Specifically, we set the other parameters to the same values as in the previous case. The feedback-related parameters are $\tau=2\pi/\omega_0$, $L = \tau c/2$, and $G_0 = \sqrt{2c\gamma/\pi}$. We then calculate $g^{(2)}(0)$ and the concurrence by scanning $\Delta$ from $-1500\kappa$ to $1500\kappa$. We vary $\Delta$ and observe $g^{(2)}(0)$. The function reaches maxima of $4.06$ at $-640\kappa$ and $640\kappa$, and maxima of $15.3$ at $-1260\kappa$ and $1260\kappa$. It reaches minima of $0.30$ at $-960\kappa$ and $960\kappa$, and a minimum of $1.035$ at $360\kappa$ and $-360\kappa$, as shown by the red curve in Fig.~\ref{Delta}(a). The plot of $g^{(2)}(0)$ versus $\Delta$ shows pronounced, asymmetric peaks (e.g., at $\Delta \approx \pm 1260\kappa$, $g^{(2)}(0)\approx 15.3$) and deep troughs (e.g., at $\Delta \approx \pm 960\kappa$, $g^{(2)}(0)\approx 0.30$). The peaks indicate strong photon bunching, signifying enhanced two-photon processes. The troughs below $1$ indicate strong photon antibunching, a signature of photon blockade. Here, $g^{(2)}(0)$ is used to observe strong antibunching (down to $0.30$) at specific detunings $\Delta\approx\pm960\kappa$, which is a signature of unconventional photon blockade driven by feedback-induced quantum interference. The feedback, combined with specific detunings, resonantly enhances transitions to either single-excitation or multi-excitation manifold states, dramatically amplifying the native nonlinear effects of the hybrid system.

Furthermore, the concurrence between the magnon and the two-level atom is computed; it peaks at $7.05\times10^{-11}$ for points $\pm 1240\kappa$, and at a higher value of $9.5\times10^{-11}$ for points $\pm 640\kappa$. Conversely, it drops to minima of $2.096\times10^{-11}$ at points $\pm 1020\kappa$ and $7.99\times10^{-11}$ at points $\pm 480\kappa$, as shown in Fig.~\ref{Delta}(b). The concurrence versus $\Delta$ shows that entanglement peaks of $9.5\times10^{-11}$ at $\Delta\approx\pm640\kappa$ occur at detunings distinct from, but related to, the extrema in $g^{(2)}(0)$. This demonstrates that maximal entanglement does not necessarily coincide with extreme photon statistics. The feedback and detuning can be tuned to prioritize the creation of specific entangled states over purely statistical features of the light field.

To compare with the feedback case, we investigate the system's behavior in the ``no feedback'' baseline as a crucial control numerical experiment. The feedback distance is set to zero ($L=0$). Other parameters are set to the same values as in the feedback case. We then calculate the concurrence and $g^{(2)}(0)$. The results are shown by the black dash-dotted curves in Fig.~\ref{Delta}. Compared with the feedback case, both $g^{(2)}(0)$ and the concurrence show significantly muted responses. Without the coherent feedback loop ($L=0$), the system lacks reinforced interference pathways. The extreme peaks of bunching/antibunching and the enhanced entanglement peaks vanish or are greatly reduced. This direct comparison definitively proves that the dramatic effects observed are due to engineered feedback, not just intrinsic system properties. The result highlights the active enhancement capability of feedback control.

\subsection{Effect of drive $\Omega_e$}

\begin{figure}[htbp]
\centering
\includegraphics[width=1\linewidth]{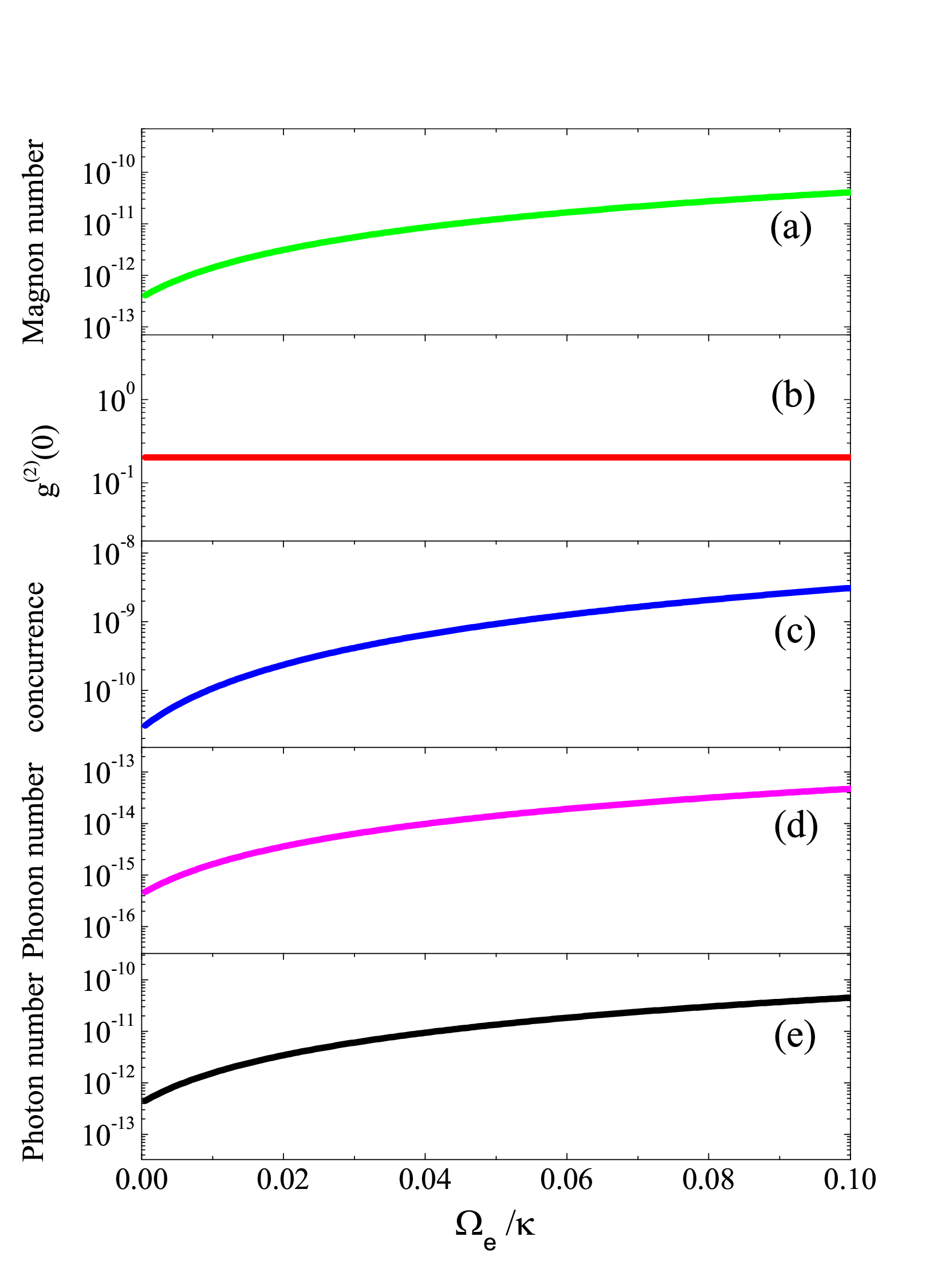}
\caption{(Color online) Behavior of (a) magnon number, (b) the second-order correlation function $g^{(2)}(0)$, (c) concurrence, (d) phonon number, and (e) photon number as functions of the pump strength $\Omega_e/\kappa$. By increasing the drive, antibunching remains constant, while the magnon number, concurrence, phonon number, and photon number increase.}
\label{fig:Omegae}
\end{figure}

With an increase in the drive $\Omega_e$, the photon number increases, as shown in Fig.~\ref{fig:Omegae}(e). This leads to an increase in the magnon number, as shown in Fig.~\ref{fig:Omegae}(a), and also to an increase in the phonon number, because the cavity is connected to the optomechanical crystal, as shown in Fig.~\ref{fig:Omegae}(d). The concurrence between the magnon and the two-level atom increases because the photon density inside the cavity increases, and it exchanges energy with the magnon and atom more frequently. Therefore, the concurrence increases, as shown in Fig.~\ref{fig:Omegae}(c).

$\Omega_e$ varies from $0.0005\kappa$ to $0.1\kappa$. In this weak-drive regime, $g^{(2)}(0)$ remains constant at $\approx0.203$ to within numerical accuracy, confirming that the normalized photon statistics are independent of the drive strength, as shown in Fig.~\ref{fig:Omegae}(b). The concurrence scales as $\Omega_e^2$ (slope 2 on a log-log plot), as do the average photon, magnon, and phonon numbers. This scaling is a direct consequence of perturbation theory: the wavefunction is $|\psi\rangle \approx |0\rangle + \Omega_e |\psi^{(1)}\rangle + \Omega_e^2 |\psi^{(2)}\rangle$, so the single-excitation amplitudes are $\propto \Omega_e$, the double-excitation amplitudes are $\propto \Omega_e^2$, leading to $g^{(2)}(0) \propto \Omega_e^4/(\Omega_e^2)^2 = \text{constant}$.

\subsection{Effect of effective photon-magnon-phonon coupling strength $\eta$}

\begin{figure}[htbp]
\centering
\includegraphics[width=1\linewidth]{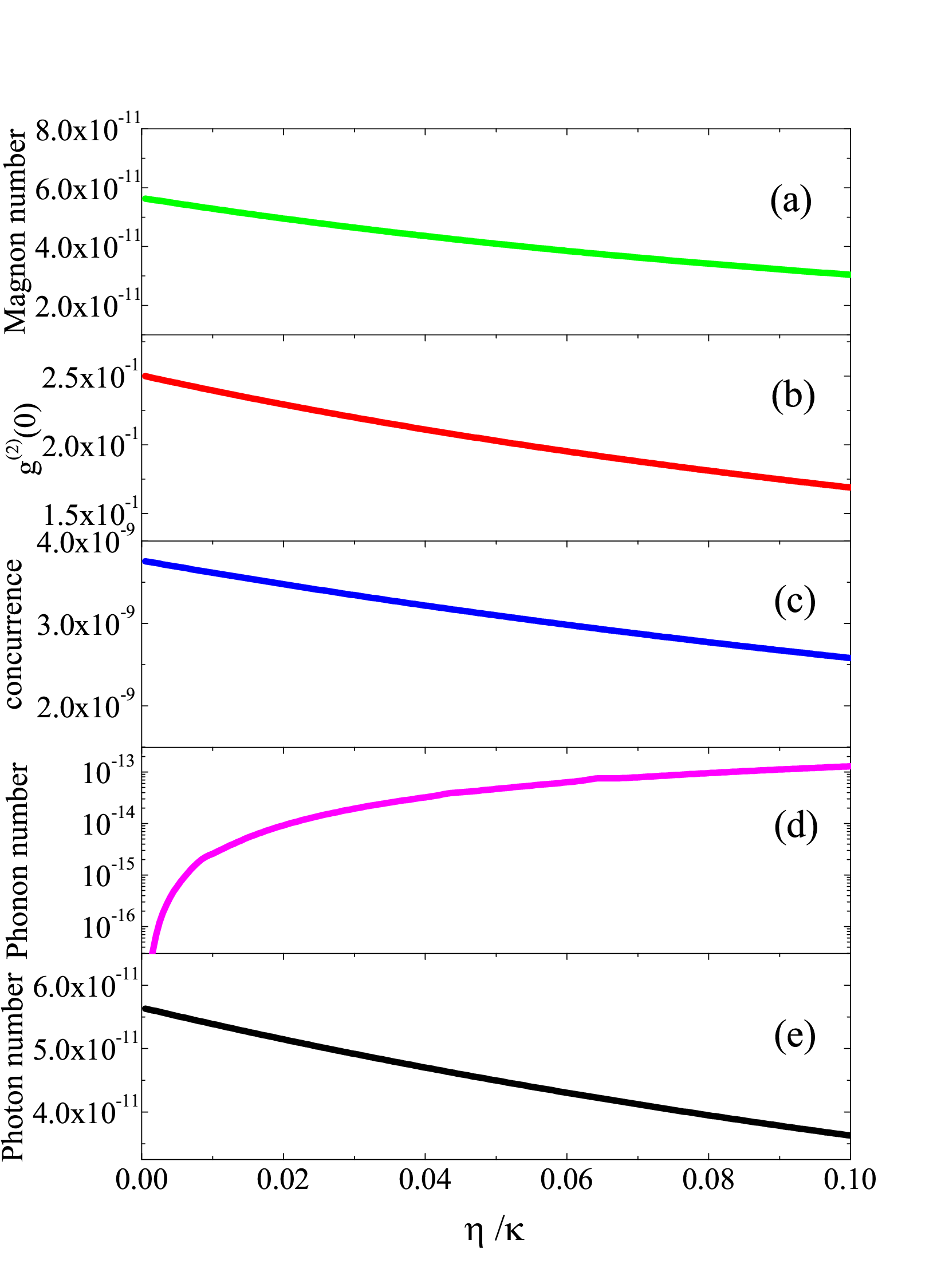}
\caption{Impact of the parameter $\eta/\kappa$ on the system. We consider (a) magnon number ($\langle m^\dagger m\rangle$), (b) photon–photon correlation function $g^{(2)}(0)$, (c) concurrence, (d) phonon number ($\langle b^\dagger b\rangle$), and (e) photon number ($\langle a^\dagger a\rangle$). All values decrease as $\eta/\kappa$ increases except the phonon number.}
\label{fig:eta}
\end{figure}

Fig.~\ref{fig:eta} displays the results for $\eta$ scanned from $0$ to $0.1\kappa$. As $\eta$ increases, $g^{(2)}(0)$ decreases monotonically from $0.250$ (at $\eta=0$) to $0.169$ (at $\eta=0.1\kappa$). This indicates that phonon-mediated three-wave mixing provides an additional decay channel for two-photon states, thereby enhancing the photon blockade effect, as shown in Fig.~\ref{fig:eta}(b). The concurrence also decreases with $\eta$, because quantum resources are shared among more degrees of freedom (monogamy of entanglement), as shown in Fig.~\ref{fig:eta}(c). The phonon number increases from zero at $\eta=0$ (no coupling) to about $1.3\times10^{-13}$ at $\eta=0.1\kappa$, as shown in Fig.~\ref{fig:eta}(d). Although the photon and magnon numbers decrease by about 30\%, consistent with the idea that part of the excitation energy is transferred to the phonon mode, as shown in Fig.~\ref{fig:eta}(e) and Fig.~\ref{fig:eta}(a).

\subsection{Effect of feedback strength $G_0$}

\begin{figure}[htbp]
\centering
\includegraphics[width=1\linewidth]{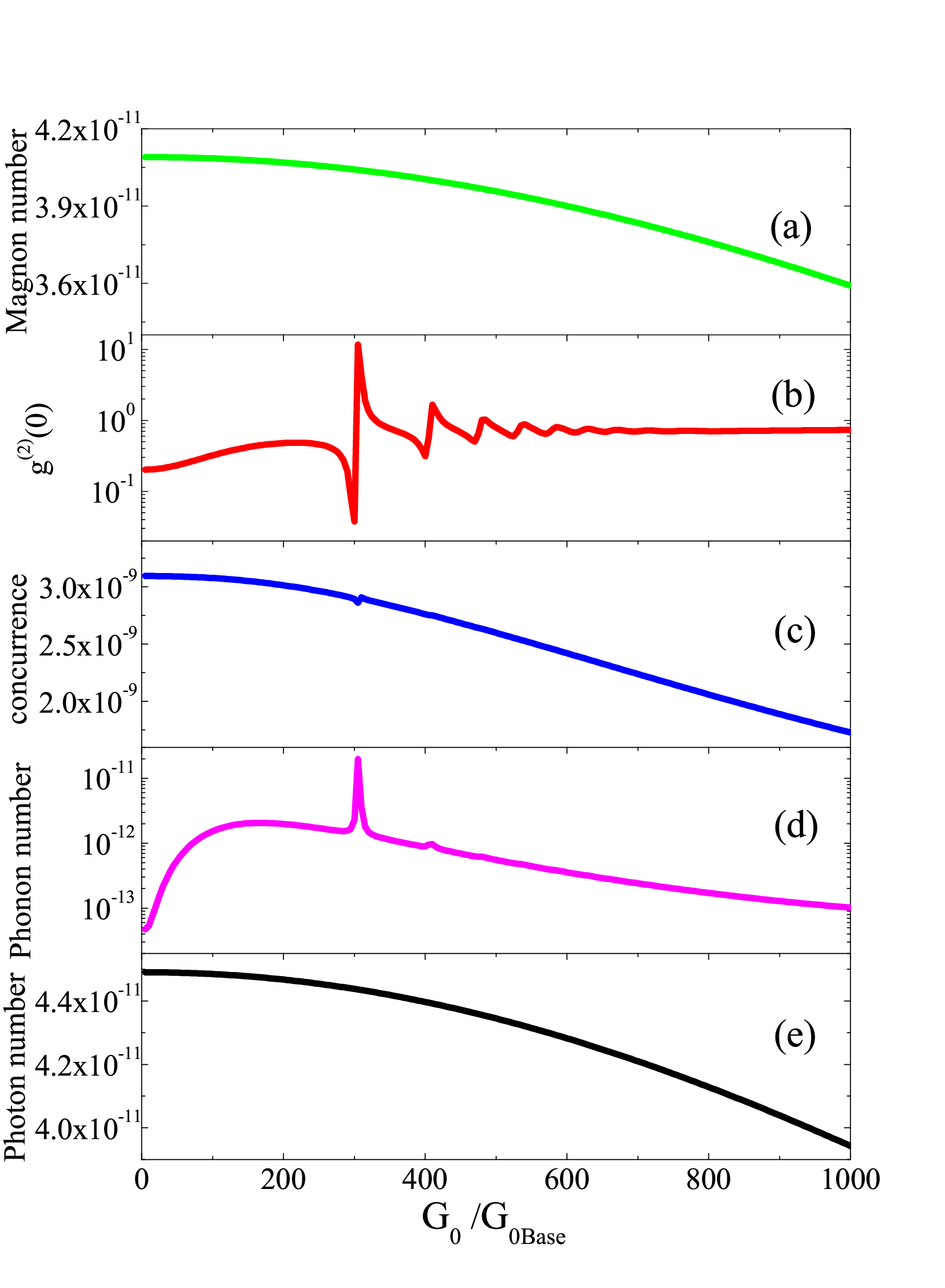}
\caption{Behavior of (a) the magnon number, (b) $g^{(2)}(0)$, (c) concurrence, (d) phonon number, and (e) photon number as functions of the coupling strength $G_0/G_{0\mathrm{Base}}$ between the cavity and the waveguide.}
\label{fig:G0}
\end{figure}

The most dramatic effects are observed when the feedback coupling $G_0$ varies from $0$ to $1000\,G_{0\mathrm{Base}}$, where $G_{0\mathrm{Base}} = \sqrt{2c\kappa/\pi}$. The results are summarized in Fig.~\ref{fig:G0}.

The magnon number [Fig.~\ref{fig:G0}(a)] and photon number [Fig.~\ref{fig:G0}(e)] decrease monotonically with $G_0$, and their ratio remains constant at $\langle m^\dagger m\rangle / \langle a^\dagger a\rangle \approx 0.9108$. This is because the waveguide couples symmetrically to the two polariton modes $a_+$ and $a_-$, so the relative population is preserved.

The value of $g^{(2)}(0)$ starts at $0.203$ for $G_0=0$, then decreases slightly. At $G_0/G_{0\mathrm{Base}}\approx 302$, it jumps to a large peak of $\sim11.7$, indicating strong photon bunching. For even larger $G_0$, it drops back to $\sim0.74$ (slight antibunching). This nonmonotonic behavior arises from quantum interference between different emission paths: the direct two-photon emission and the path where one photon leaves, returns after a delay, and then participates in a second emission. At the resonant $G_0$, constructive interference dramatically enhances the two-photon probability, as shown in Fig.~\ref{fig:G0}(b).

The concurrence decreases with $G_0$, but at the same resonance it exhibits a small upward jump, as shown in Fig.~\ref{fig:G0}(c). This reflects a temporary redistribution of quantum resources from the waveguide to the atom–magnon subsystem when the two-photon resonance condition is met.

The phonon number [Fig.~\ref{fig:G0}(d)] shows a spectacular enhancement at the resonance, rising from the background level $\sim4.7\times10^{-14}$ to $\sim1.98\times10^{-11}$, i.e., an enhancement factor of about $423$. This is because the resonantly enhanced two-polariton state decays efficiently via the three-wave mixing process $a_+ \to a_- + b$, depositing energy into the phonon mode.

\section*{\label{sec:level1} Experimental feasibility discussion}

The parameter values, though numerically idealized, correspond to physical mechanisms that are realized in modern quantum optics experiments, most typically in microwave-frequency cavity magnonic systems. The following is a mapping of the key parameters to the experimental platforms.

First, magnon–photon coupling is strong ($G_m = 200\kappa$, $g = 40\gamma$) \cite{Goryachev2014}. This requires the coupling rate to greatly exceed all loss rates \cite{Bourhill2016}. A typical platform consists of a high-purity YIG sphere inside a high-$Q$ three-dimensional microwave cavity \cite{Tabuchi2014}. The experimental coupling strength $G_m$ between the magnon and the photon can be as high as 2\,GHz or even 7.11\,GHz, far exceeding cavity losses \cite{Goryachev2014}.

Second, phonon–photon coupling can be realized experimentally as well. In our model, the phonon frequency is set to $\omega_b = 2G_m$ and $g = 40\gamma$. $\omega_b$ is on the order of gigahertz, similar to the magnon–photon coupling strength, enabling coherent energy exchange. This has been realized using silicon optomechanical crystals \cite{Chan2011PRL}. The experiment demonstrated that the mechanical frequencies (phonon modes) are around 2.5\,GHz, with vacuum optomechanical coupling rates in the MHz range \cite{Chan2011PRL}. It also achieved ground-state cooling for the first time. At the same time, mechanical frequencies of $5.91\,\text{GHz}$ and $g/2\pi = 1.9\,\text{MHz}$ have been realized in experiments \cite{Chan2011Nature}. Later, dimethylsilane optomechanical crystals \cite{Matheny2018} realized $g/2\pi = 1.70\,\text{MHz}$ with a mechanical frequency of $1.14\,\text{GHz}$. Furthermore, a platform of two-dimensional optomechanical crystals has been realized with a mechanical frequency of $7.4\,\text{GHz}$ \cite{Mayor2025} and $g/2\pi \approx 880\,\text{kHz}$. This is compatible with low temperatures. Therefore, with $\omega_b = 2G_m$, the mechanical frequency $\omega_b$ in these experiments lies in the interval from $1\,\text{GHz}$ to $10\,\text{GHz}$, while the photon–phonon coupling rate ranges from $0.9\,\text{MHz}$ to $1.9\,\text{MHz}$. These parameters match the values applied in our model.

Third, given $\gamma = \kappa$, all parameters take $\kappa = 5\times10^{-6}\times 0.4\pi\,\text{THz} = 2\pi\times 1\,\text{MHz}$ as a benchmark. In experiments, a superconducting cavity at millikelvin temperatures \cite{Tabuchi2014} realizes $\kappa/2\pi$ on the order of MHz, and the cavity photon lifetime is sufficiently long, thus satisfying the requirements for coherent dynamics. At the same time, the low-temperature condition confirms the coherence of magnons and enables ground-state cooling of the optomechanical crystal. In addition, the low-temperature environment suppresses thermal noise and prolongs the coherence time.

Fourth, weak driving of $\Omega_e = 0.1\kappa$ and low excitation justify the few-excitation subspace truncation. Cavity QED or cavity magnonics in the weak-field regime is easy to realize. Experimentally, with a weak microwave drive, single-excitation probabilities on the order of $10^{-9}$ are entirely realistic.

Lastly, our model uses time-delayed coherent feedback ($G$, $L$, $\tau$) through a waveguide and mirror. The same feedback-induced quantum interference scheme has been experimentally verified in atomic ensembles, showing that varying the feedback delay ($L$) can switch and enhance photon statistics.

In short: (i) YIG sphere microwave cavities achieve strong magnon–photon coupling (matching $G_m$). (ii) Cryogenic temperatures suppress thermal noise and ensure long coherence. (iii) Optomechanical crystals realize phonon–photon coupling via radiation pressure or the photoelastic effect. (iv) The feedback loop can be realized by a mirror in a waveguide, as demonstrated in follow-up experiments. Thus, the parameter regime of this paper is fully achievable in microwave-frequency cavity magnomechanical systems.

\section*{\label{sec:level1} Conclusion and Outlook}

Coherent feedback control is a powerful method for actively tailoring and enhancing the quantum statistical properties and correlations in complex hybrid quantum systems. First, feedback is a potent, tunable controller: the distance $L$ acts as a periodic phase knob, enabling real-time switching of the system's quantum output. Second, feedback amplifies nonlinearity: when combined with detuning, it drives the system into regimes of extreme quantum statistics (strong antibunching/bunching) that are far more pronounced than those in the passive system. Third, feedback directly manipulates entanglement: it enhances matter-based quantum correlations (concurrence) by stabilizing specific entangled states. Finally, the link between statistics and entanglement is tunable: feedback allows one to navigate the relationship between photon statistics ($g^{(2)}$) and entanglement (concurrence), enabling optimization for different quantum information tasks (e.g., single-photon generation versus entangled pair creation). In this work, we have introduced a magnon into the feedback loop for the first time and employed an LSTM network—a machine learning approach—to efficiently propagate the probability amplitudes to the steady state, confirming that photon statistics are not merely output indicators but are deeply intertwined with the internal quantum state of matter. Future directions may involve integrating feedback control into multi-mode hybrid systems to stabilize quantum correlations against decoherence, or using the photon-statistics–entanglement link for non-destructive probing of complex quantum states. These approaches are vital for advancing quantum technologies such as on-demand single-photon sources, quantum memories, and interconnected quantum networks.

\section*{Acknowledgements}Y. L. would like to thank Haoen Feng, Yuan Yao, Feiyang Liu and Yuhan Dai for the valuable discussions. Y.L. is supported by Start-up Funding for Research from China Jiliang University. S.L. acknowledges the funding from National Natural Science Foundation of China (Grant No. 12104296)

\section*{Appendix 1}

The equations of motion for the time evolution of the state are given by the Schrödinger equation. Since weak driving is considered, the coefficient $|C_{g0000}|$ is approximated as unity, so that the equations of motion become

\begin{equation}\label{EqMotion}
\begin{aligned}
\frac{\partial C_{g1000}}{\partial t} &= -i \biggl( \Delta_w C_{g1000} - \eta C_{g0110} + \eta_a C_{e0000} + \Omega_e C_{e1000} \\
&\qquad + \int G_0 \sin(k L) \exp[i t (w_l - w_k)] C_{g000k} \, dk \biggr), \\
\frac{\partial C_{g2000}}{\partial t} &= -i \biggl( 2\Delta_w C_{g2000} - \sqrt{2}\eta C_{g1110} + \sqrt{2}\eta_a C_{e1000} \\
&\qquad + \int G_0 \sin(k L) \exp[i t (w_l - w_k)] C_{g100k} \, dk \biggr), \\
\frac{\partial C_{g1110}}{\partial t} &= -i \biggl( (2\Delta_w - 2G_m + w_b) C_{g1110} - 2\eta C_{g0220} \\
&\qquad - \sqrt{2}\eta C_{g2000} + \eta_a C_{e0110} \\
&\qquad + \int G_0 \sin(k L) \exp[i t (w_l - w_k)] C_{g011k} \, dk \biggr), \\
\frac{\partial C_{e1000}}{\partial t} &= -i \biggl( 2\Delta_w C_{e1000} - \eta C_{e0110} + \sqrt{2}\eta_a C_{g2000} + \Omega_e C_{g1000} \\
&\qquad + \int G_0 \sin(k L) \exp[i t (w_l - w_k)] C_{e000k} \, dk \biggr), \\
\frac{\partial C_{g000kk}}{\partial t} &= -i \sqrt{2} \int G_0 \sin(k L) \exp[i t (w_l - w_k)] C_{g100k} \, dk, \\
\frac{\partial C_{g000k}}{\partial t} &= -i \int G_0 \sin(k L) \exp[-i t (w_l - w_k)] C_{g1000} \, dk, \\
\frac{\partial C_{g011k}}{\partial t} &= -i \biggl( ((\Delta_w - 2G_m) + w_b) C_{g011k} - \eta C_{g100k} \\
&\qquad + \int G_0 \sin(k L) \exp[-i t (w_l - w_k)] C_{g1110} \, dk \biggr), \\
\frac{\partial C_{g0110}}{\partial t} &= -i \bigl( ((\Delta_w - 2G_m) + w_b) C_{g0110} - \eta C_{g1000} + \Omega_e C_{e0110} \bigr), \\
\frac{\partial C_{e0000}}{\partial t} &= -i \bigl( \Delta_w C_{e0000} + \eta_a C_{g1000} + \Omega_e C_{g0000} \bigr), \\
\frac{\partial C_{g0220}}{\partial t} &= -i \bigl( (2(\Delta_w - 2G_m) + 2w_b) C_{g0220} - 2\eta C_{g1110} \bigr), \\
\frac{\partial C_{e0110}}{\partial t} &= -i \bigl( (\Delta_w - 2G_m + w_b) C_{e0110} + \Delta_w C_{e0110} \\
&\qquad - \eta C_{e1000} + \eta_a C_{g1110} + \Omega_e C_{g0110} \bigr)\\
\frac{\partial C_{g100k}}{\partial t} &= -i \biggl( \Delta_w C_{g100k} - \eta C_{g011k} + \eta_a C_{e000k} \\
&\qquad + \int \sqrt{2} G_0 \sin(k L) \exp[i t (w_l - w_k)] C_{g2000} \, dk \\
&\qquad + \int_{-\infty}^{k-} G_0 \sin(k L) \exp[i t (w_l - w_k)] C_{g000kk'} \, dk \\
&\qquad + \int_{k+}^{+\infty} G_0 \sin(k L) \exp[i t (w_l - w_k)] C_{g000kk'} \, dk \\
&\qquad + \sqrt{2} G_0 \sin(k L) \exp[i t (w_l - w_k)] C_{g000kk} \, dk \biggr).\\
\frac{\partial C_{g000kk'}}{\partial t} &= -i \biggl( \int G_0 \sin(k L) \exp[-i t (w_l - w_k)] C_{g100k} \, dk \biggr),\nonumber
\end{aligned}
\end{equation}

\begin{equation}
\begin{aligned}
\frac{\partial C_{e000k}}{\partial t} &= -i \biggl( \Delta_w C_{e000k} + \eta_a C_{g100k} + \Omega_e C_{g000k} \\
&\qquad + \int G_0 \sin(k L) \exp[i t (w_l - w_k)] C_{e1000} \, dk \biggr), 
\end{aligned}
\end{equation}

\section*{Appendix 2}
 Since the system is in the weak driving regime, we assume $C_{g0000} = 1$. When steady state is achieved, the partial derivative of the probability amplitude of the state is zero. Therefore, we have
 
\begin{equation}
\begin{aligned}
0 &= -i \biggl( \Delta_w C_{g1000} - \eta C_{g0110} + \eta_a C_{e0000} + \Omega_e C_{e1000} \\
&\qquad + \int G_0 \sin(kL) \exp[i t (w_l - w_k)] C_{g000k} \, dk \biggr), \\
0 &= -i \biggl( 2\Delta_w C_{g2000} - \sqrt{2}\,\eta C_{g1110} + \sqrt{2}\,\eta_a C_{e1000} \\
&\qquad + \int G_0 \sin(kL) \exp[i t (w_l - w_k)] C_{g100k} \, dk \biggr), \\
0 &= -i \biggl( (2\Delta_w - 2G_m + w_b) C_{g1110} - 2\eta C_{g0220} \\
&\qquad - \sqrt{2}\,\eta C_{g2000} + \eta_a C_{e0110} \\
&\qquad + \int G_0 \sin(kL) \exp[i t (w_l - w_k)] C_{g011k} \, dk \biggr), \\
0 &= -i \biggl( 2\Delta_w C_{e1000} - \eta C_{e0110} + \sqrt{2}\,\eta_a C_{g2000} + \Omega_e C_{g1000} \\
&\qquad + \int G_0 \sin(kL) \exp[i t (w_l - w_k)] C_{e000k} \, dk \biggr), \\
0 &= -i \sqrt{2} \int G_0 \sin(kL) \exp[i t (w_l - w_k)] C_{g100k} \, dk, \\
0 &= -i \biggl( \Delta_w C_{g100k} - \eta C_{g011k} + \eta_a C_{e000k} \\
&\qquad + \int \sqrt{2}\, G_0 \sin(kL) \exp[i t (w_l - w_k)] C_{g2000} \, dk \\
&\qquad + \int_{-\infty}^{k-} G_0 \sin(kL) \exp[i t (w_l - w_k)] C_{g000kk'} \, dk \\
&\qquad + \int_{k+}^{+\infty} G_0 \sin(kL) \exp[i t (w_l - w_k)] C_{g000kk'} \, dk \\
&\qquad + \sqrt{2}\, G_0 \sin(kL) \exp[i t (w_l - w_k)] C_{g000kk} \, dk \biggr),\\
0 &= -i \int G_0 \sin(kL) \exp[-i t (w_l - w_k)] C_{g1000} \, dk, \nonumber
\end{aligned}
\end{equation}
\begin{equation}
\begin{aligned}
0 &= -i \biggl( ((\Delta_w - 2G_m) + w_b) C_{g011k} - \eta C_{g100k} \\
&\qquad + \int G_0 \sin(kL) \exp[-i t (w_l - w_k)] C_{g1110} \, dk \biggr), \\
0 &= -i \bigl( ((\Delta_w - 2G_m) + w_b) C_{g0110} - \eta C_{g1000} + \Omega_e C_{e0110} \bigr),\\
0 &= -i \bigl( \Delta_w C_{e0000} + \eta_a C_{g1000} + \Omega_e C_{g0000} \bigr)\\
0 &= -i \bigl( (\Delta_w - 2G_m + w_b) C_{e0110} + \Delta_w C_{e0110} \\
&\qquad - \eta C_{e1000} + \eta_a C_{g1110} + \Omega_e C_{g0110} \bigr),\nonumber
\end{aligned}
\end{equation}
\begin{equation}
\begin{aligned}
0 &= -i \bigl( (2(\Delta_w - 2G_m) + 2w_b) C_{g0220} - 2\eta C_{g1110} \bigr)\nonumber
\end{aligned}
\end{equation}
\begin{equation}
\begin{aligned}
0 &= -i \biggl( \Delta_w C_{e000k} + \eta_a C_{g100k} + \Omega_e C_{g000k} \\
&\qquad + \int G_0 \sin(kL) \exp[i t (w_l - w_k)] C_{e1000} \, dk \biggr), \\
0 &= -i \biggl( \int G_0 \sin(kL) \exp[-i t (w_l - w_k)] C_{g100k} \, dk \biggr).
\end{aligned}
\end{equation}\label{EqMotionSteady}

\end{document}